\documentclass[]{interact}
\usepackage{xcolor}
\usepackage{booktabs}
\usepackage{makecell}
\usepackage{siunitx}
\usepackage{epstopdf}
\usepackage[caption=false]{subfig}
\usepackage{hyperref}
\usepackage[longnamesfirst,sort]{natbib}
\bibpunct[, ]{(}{)}{;}{a}{,}{,}
\renewcommand\bibfont{\fontsize{10}{12}\selectfont}

\theoremstyle{plain}

\theoremstyle{definition}

\theoremstyle{remark}

\begin{document}


\articletype{\textbf{PREPRINT}}


\title{Impact of UK Postgraduate Student Experiences on Academic Performance in Blended Learning: A Data Analytics Approach}



\author{
\textbf{\name{Muhidin Mohamed\textsuperscript{*}\thanks{*Email: m.mohamed10@aston.ac.uk}, Shubhadeep Mukherjee and Bhavana Baad}}
\affil{Dept. of Business Analytics and Information Systems, Aston University, Birmingham, UK}
}
\maketitle
 \vspace{-10pt}
\begin{abstract}
Blended learning has become a dominant educational model in higher education in the UK and worldwide, particularly after the COVID-19 pandemic. This is further enriched with accompanying pedagogical changes, such as strengthened asynchronous learning, and the use of AI (from ChatGPT and all other similar tools that followed) and other technologies to aid learning. While these educational transformations have enabled flexibility in 
learning and resource access, they have also exposed new challenges on how students can construct successful learning in hybrid learning environments. In this paper, we investigate the interaction between different dimensions of student learning experiences (ranging from perceived acceptance of teaching methods and staff support/feedback to learning pressure and student motivation) and academic achievement within the context of postgraduate blended learning in UK universities. To achieve this, we employed a combination of several data analytics techniques including visualization, statistical tests, regression analysis, and latent profile analysis. Our empirical results (based on a survey of 255 postgraduate students and holistically interpreted via the Community of Inquiry (CoI) framework) demonstrated that learning activities combining teaching and social presences, and tailored academic support through effective feedback are critical elements for successful postgraduate experience in blended learning contexts. Regarding contributions, this research advances the understanding of student success by identifying the various ways demographic, experiential, and psychological factors impact academic outcomes. And in theoretical terms, it contributes to the extension of the CoI framework by integrating the concept of learner heterogeneity and identifying four distinct student profiles based on how they engage in the different CoI presences.

\end{abstract}


\section{Introduction}
The rapid digital and pedagogical transformation of Higher Education (HE), accelerated by the COVID-19 pandemic, has greatly reshaped the way learning is designed, delivered and experienced. For example, the widespread adoption of Blended Learning (BL) -- integrating synchronous online interactions with asynchronous activities – has made self-paced study more than an emergency solution, defining a new pedagogical norm that underpins sustainable HE learning model ~\citep{harper2024face,cheung2023online}. Yet, while this shift has enabled flexibility and access, it has also exposed new challenges in how students engage cognitively, socially, and emotionally in hybrid environments. These challenges may not be evenly distributed: students differ markedly in how they experience and benefit from blended learning settings. Understanding this heterogeneity is critical for both pedagogy and institutional strategy.

Recent educational research underscores that BL serves as an effective learning model for mitigating crisis and emergencies such as the COVID-19 pandemic, pushing for further pedagogical transformation in the post-pandemic era \citep{mohammadi2025blended,dei2024promoting,cobo2022return}. Many UK universities have retained elements of the Online Blended Learning (OBL) model introduced during the pandemic such as pre-recorded lectures and live online discussions while reintroducing in-person elements to support community and motivation~\citep{harper2024face}. This hybridization has created a new learning ecology, one characterized by simultaneous digital immediacy and reflective distance. However, with this flexibility come increased demands on learners’ self-regulation, time management, and emotional coping. Postgraduate students, in particular, are expected to navigate work commitments alongside academic expectations, requiring higher digital literacy and self-direction~\citep{cheung2023online, anthony2022blended}. As such, the blended environment functions as both an opportunity and a stressor intensifying the need to understand which learners benefit from specific modes of engagement and why.

Blended learning offers various advantages that collectively make effective education model including interaction with learning community, flexible self-paced independent learning, digital platforms for educational support and feedback, reflection on learning, etc. Yet, in practice, not all learners thrive equally. The complexity of online engagement introduces what may be described as 'uneven presence' across students: some are able to maintain motivation and self-regulation amid digital fatigue, while others struggle with isolation, stress, or cognitive overload. These differences invite fundamental questions about the Community of Inquiry (CoI) framework~\citep{garrison2001critical}, which argues that meaningful learning emerges through the interplay of teaching presence (design and facilitation of learning), social presence (affective and relational engagement), and cognitive presence (construction of meaning through reflection and inquiry). The CoI model has been extensively validated in online education, but its latent assumption of a relatively homogeneous learner experience remains under-examined~\citep{castellanos202020, carroll2025extended}.

The present study examines these dynamics empirically within the context of postgraduate blended learning in UK universities, exploring how different dimensions of the learning experience ranging from workload and stress to motivation, feedback, and online engagement among others shape academic performance. These student experience factors are considered to be significant predictors of academic performance, therefore justifying the need for empirical evidence of their associations under the online blended learning mode~\citep{prananto2025perceived}.  Using survey data from 255 postgraduate students, we employ a combination of data summarization and visualization, statistical tests, regression analysis, and Latent Profile Analysis (LPA) to uncover these interactions and underlying learner profiles that emerge within blended settings. The study's key goals are to understand which factors significantly predict academic outcomes, identify the difference between student subgroups based on learning materials, satisfaction, etc, and finally appreciate how combinations of cognitive, social, and teaching presences merge into distinct experiential patterns that influence learning success.

In theoretical terms, this research contributes to the extension of the CoI framework by integrating the concept of learner heterogeneity. Traditional CoI research assumes a unified learner experience, focusing on the instructor's facilitative role and the technological affordances that sustain engagement. Our results challenge this assumption by demonstrating that in blended postgraduate classrooms, multiple CoI may coexist each with its own equilibrium of teaching, social, and cognitive presences. This suggests that HE educators must not only design for collective engagement but also recognize micro-level differences in how students construct meaning and cope with blended demands. Moreover, this extension invites further theorization around how psychological and motivational factors interact dynamically with the CoI dimensions over time, potentially reframing the model as a differentiated and adaptive framework rather than a static triad. Practically, our findings highlight the need for targeted interventions within postgraduate blended programs. By connecting these insights to the CoI framework, institutions can move beyond one-size-fits-all models of online engagement toward a more nuanced, evidence-based understanding of how blended pedagogies shape learning outcomes.

In light of these considerations, this study enhances understanding of how various dimensions of students’ learning experiences interact with academic performance in blended learning contexts by addressing the following key questions.
\begin{enumerate}
\item In what ways do students respond to different BL methods, staff support \& feedback, and academic overload to construct successful learning?
\item How do teaching, social and cognitive presences interact to influence postgraduate students’ academic performance in BL environments?
\item To what extent do psychological and motivational mechanisms (e.g., stress, workload, motivation) mediate or moderate the relationship between BL engagement and academic performance?
\item What are the distinct learner profiles within blended postgraduate classrooms, and how do these profiles differ in their balance of the three CoI presences?
\end{enumerate}

To explore these questions, we collected a sample data through an online survey of postgraduate taught students in UK universities. The study examines variables including study workload, stress levels, teaching feedback, motivation, and live online participation as core elements of blended learning design and analyses their direct and indirect effects on self-assessed academic performance. By aligning these empirical findings with the Community of Inquiry framework, we aim to offer a theoretically grounded yet practically relevant account of how postgraduate learners navigate the complexities of blended education in a post-pandemic world. Ultimately, we argue that blended learning environments are not homogeneous ecosystems but differentiated communities of inquiry.

\section{Literature Review and the CoI Framework} 
The shift towards blended learning in higher education especially in the aftermath of the COVID-19 pandemic has significantly changed how postgraduate students engage with academic content, instructors, and peers. Defined as a hybrid approach largely combining  online live classes and asynchronous (e.g. recorded) materials, blended learning offers flexibility and personalization, making it especially appealing for postgraduate learners managing multiple commitments. This literature review critically explores the postgraduate student experience in UK universities under blended learning models, examining the influence of study load, student satisfaction, mental health, teaching support, self-regulation, and technology on academic achievement. It also discusses the study’s guiding educational model, the CoI framework.

\subsection{Related works} 
The widespread recent adoption of blended learning and its successes in higher education makes it a new pedagogical approach that could underpin sustainable HE learning for years to come. It is praised for enabling flexible learning paths and increasing accessibility for postgraduate students~\citep{cheung2023online, anthony2022blended}. Linked with this, \cite{ritz2023support} emphasize the transformative potential of blended learning when combined with effective technological interventions, especially during periods of disruption such as the COVID-19 crisis. Besides, studies such as \cite{castro2019blended} have noted that educational technology capabilities (ETCs) like video capsules and intelligent tutoring systems support personalised learning and allow students to progress at their own pace, thereby enhancing academic achievement. These technologies not only make course materials more accessible but also improve feedback mechanisms and foster autonomy. These interventions increase student engagement and foster meaningful learning outcomes, underscoring the value of integrating digital tools purposefully within academic programs. Students' academic achievement could be percieved as a function of various pedagogical factors including learner satisfaction, academic support, mental well-being, self-regulation, technology use, etc. 


\begin{itemize}
\item \textbf{Student satisfaction}:  Student satisfaction is a central driver of academic performance in blended environments. According to \cite{kok2025mediating}, both teacher and digital support play a critical role in enhancing student engagement, which mediates learner satisfaction in blended learning environments. This aligns with earlier findings by \cite{kuo2014k}, who linked satisfaction to factors such as interaction, course design, and self-regulated learning. \cite{de2022delivering} also found that students who transitioned from face-to-face to blended learning expressed mixed feelings. While many appreciated the online components, those accustomed to traditional formats experienced higher levels of frustration. Interestingly, mature postgraduate students especially those attending evening classes reported more positive perceptions of the blended model, appreciating its flexibility and resource accessibility.

\item\textbf{Tutor support}: Effective tutor support remains vital for student success in blended settings. According to \cite{mckenzie2022examination}, clarity in tutor roles and strong coordination in team-teaching environments contribute significantly to the success of blended learning models. They emphasize that support systems must be transparent and technologically integrated to avoid miscommunication and disengagement. This builds upon earlier frameworks by \cite{salmon2000computer} and is echoed by \cite{stone2019olders}, who note that meaningful feedback and proactive engagement by instructors reduce feelings of isolation and increase academic confidence. In the UK context, timely and personalized support is especially important for international and part-time postgraduate students.

\item\textbf{Study workload and stress}: Scholars have found workload and stress to be persistent issues in postgraduate education~\citep{richardson2015cosmopolitan,robotham2008stress}. Research has highlighted that balancing self-paced learning with in-person deadlines can lead to cognitive overload, anxiety, and burnout~\citep{beiter2015prevalence}. Related to this, \cite{de2022delivering} found that workload management becomes even more critical when transitioning students between learning modalities. Supporting students with clearer expectations and realistic pacing is essential to reduce stress and improve academic outcomes. Also, \cite{hughes2019university} found that postgraduate students in UK institutions experience higher levels of anxiety and depression, often exacerbated by unclear course structures or inconsistent support. A study by \cite{aristovnik2020impacts} reveals that during COVID-19, students’ psychological well-being was heavily influenced by the perceived quality of digital interaction and institutional responsiveness. Universities adopting well-integrated blended models with emotional check-ins and peer support structures have seen more positive outcomes.

\item \textbf{Self-regulated learning and time management}: Blended learning places increased responsibility on students to manage their own time, set goals, and stay motivated, skills collectively referred to as self-regulated learning. In this regard, \cite{zimmerman2002becoming} posited that self-regulatory capabilities strongly predict academic achievement, especially in autonomous learning environments. \cite{ritz2023support} further demonstrated that technology-enhanced learning environments can actively scaffold self-regulated learning. Through tools such as goal-setting dashboards and embedded prompts, students can better manage their progress and maintain engagement. \cite{broadbent2015self} similarly found that time management and self-monitoring were among the strongest predictors of success in online and blended learning environments.

\item\textbf{Role of technology in learning experience}: Effective integration of LMS platforms, asynchronous content (e.g., video lectures), and communication tools can significantly improve access, personalization, and engagement. \cite{castro2019blended} and \cite{kok2025mediating} both emphasize the importance of aligning technological tools with pedagogical goals to ensure a cohesive learning experience. However, the quality and consistency of digital tools matter. \cite{buhl2023scoping} call for deeper research into the face-to-face components of blended learning, arguing that we still lack comprehensive models for how in-person and digital elements interact meaningfully. Technological failures, poor design, or lack of training can exacerbate existing barriers to engagement, particularly for students unfamiliar with digital platforms \citep{kebritchi2017issues}. Improvements in educational technology enabled online synchronous activities to provide similar learning and interaction opportunities as  face2face sessions~\citep{heilporn2021examination}
\end{itemize}

Extant literature strongly emphasises that blended learning is here to stay and is constantly evolving as a defining feature of education in the UK~\citep{hill2023visions,harper2024face}. Evidence also suggests that its success relies on an intricate balance of technology integration, tutor engagement, student satisfaction, mental health support, and workload management. Despite this, we have observed notable gaps in the current literature. While studies such as \cite{castro2019blended}, \cite{mckenzie2022examination}, and \cite{de2022delivering} have advanced our understanding of blended learning components and student perceptions, there is limited empirical evidence on how these dimensions interact to influence academic achievement among UK postgraduate students. Furthermore, research has not adequately explored the unique needs of mature, international, and part-time students who navigate blended models in post-pandemic university settings. Recent work by \cite{buhl2023scoping} highlights the lack of rich characterization of in-person elements  in a blended learning environment. \cite{kok2025mediating} also call for deeper examination of engagement as a mediating factor. These gaps underscore the need for context-specific, holistic studies that examine how tutor support, workload, satisfaction, and emotional wellbeing combine to shape the academic experience under blended learning models. We try to address these questions in our paper through statistical and empirically examination of post graduate student experience in the UK, emphasising on the Community of Inquiry framework.

 
\subsection{The Community of Inquiry (CoI) framework}

The Community of Inquiry (CoI) developed by~\citep{garrison2001critical} is a framework for identifying and  understanding how various educational factors interplay to form successful learning  experience in online and blended environments in higher education. The model is built on the educational principles that a meaningful learning experience is constructed through the interaction of three core elements: teaching, cognitive, and social presences~\citep{garrison2001critical,garrison2007researching}. The teaching presence is the CoI component for designing, facilitating, and supporting learners to construct successful educational experience. Social presence, on the other hand, represents the communication, collaboration and connection of learners with the educational community in online blended learning contexts. The third element, cognitive presence, evaluates the depth of learners’ understanding as shaped by sustained discourse and collaborative reflection within the learning community. The CoI model has been widely adopted  in online and blended environments to enhance learning through effective design, facilitation, interaction, and reflection~\citep{stenbom2018systematic,castellanos202020, carroll2025extended}.


\begin{figure}
\centering
\includegraphics[width=0.65\columnwidth]{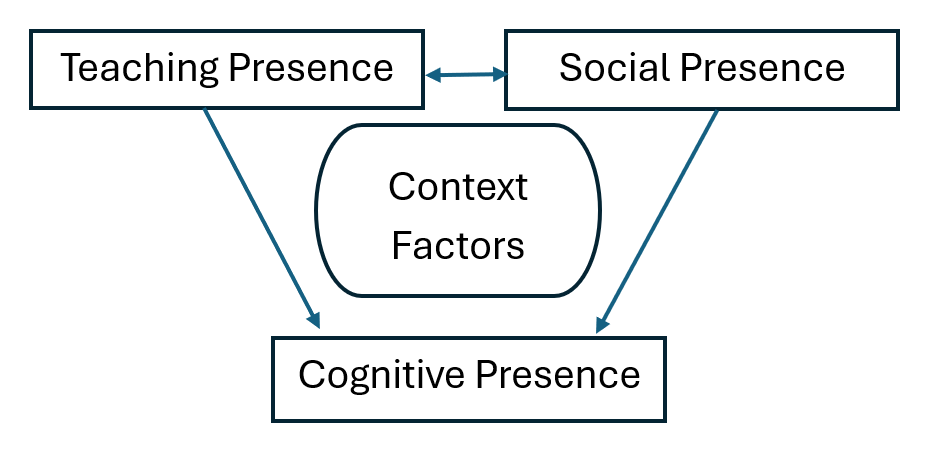}
 \caption{A high-level illustration of the Community of Inquiry framework in the study context}
    \label{CoI-mapping}
\end{figure}

 \autoref{CoI-mapping} shows the contextualization of our emprical data in the  CoI framework, indicating the three aforementioned presences and context factors (background attributes that do not belong to any of the CoI dimensions but useful for interpretting the interaction of the framework components, e.g. age, gender, etc). In the context of this study, the presence of teaching is captured through different (a)synchronous learning methods, such as recorded lecture videos -- which direct the learning and support students to acquire the intended conceptual knowledge -- and accompanying study resources that include lecture slides, hands-on exercises, etc -- which help students consolidate the learning and apply acquired concepts / skills to form functioning knowledge. In addition, educational activities such as live online classes and discussion forums were designed to facilitate social presence. Finally, we used indicators such as the self-assessed academic performance of the learners to measure the cognitive presence and the learning achievement (cf.~\autoref{tab:CoI-mapping}). The cognitive presence is considered the CoI element that captures the focus and success of the learning experience.

\section{Research objectives} 
Our study explores how various instructional, support, behavioural, and well-being factors influence postgraduate student learning experience and academic outcomes in a blended learning environment. We focus primarily on student satisfaction with teaching practices, academic support and feedback systems, perceived workkolad and stress levels, and self-reported motivation. Demographic characteristics, such as gender, age, and average study time are also considered for their relationship with learning behaviours. Specifically, we investigate the following overarching research objectives: 
\begin{enumerate}
\item Explore characteristics of study participants to understand demographic (e.g. age, gender) and behavioural (e.g. study time) patterns and their association with  academic achievement. 
\item  Assess associations between student satisfaction with various academic supports and their academic performance.
\item  Examine the influence of academic workload and stress on performance outcomes.
\item Model the joint effects of student motivation, workload, and mental state on academic attaiment.
\item Interpret study's findings through the CoI framework to identify relative contribution of teachig, social, and cognitive presence to successful learning.
\end{enumerate}

Our research has used a quantitative approach to achieve the objectives. Especially, we employed data analysis (statical summaries, visualizations, etc), statistical tests (Spearman, and Mann-Whitney U tests) to investigate objectives 1- 3, followed with empirical modelling (i.e. ordinal regression) to explore objective 4. Then, the results obtained from these analytics examinations helped us to form a holistic understanding of the student educational experience in a blended learning environment and interpret the research findings through the established CoI education model. 

\section{Research Design and Methodology}
\begin{figure}[h!]
\centering
\includegraphics[width=0.85\columnwidth]{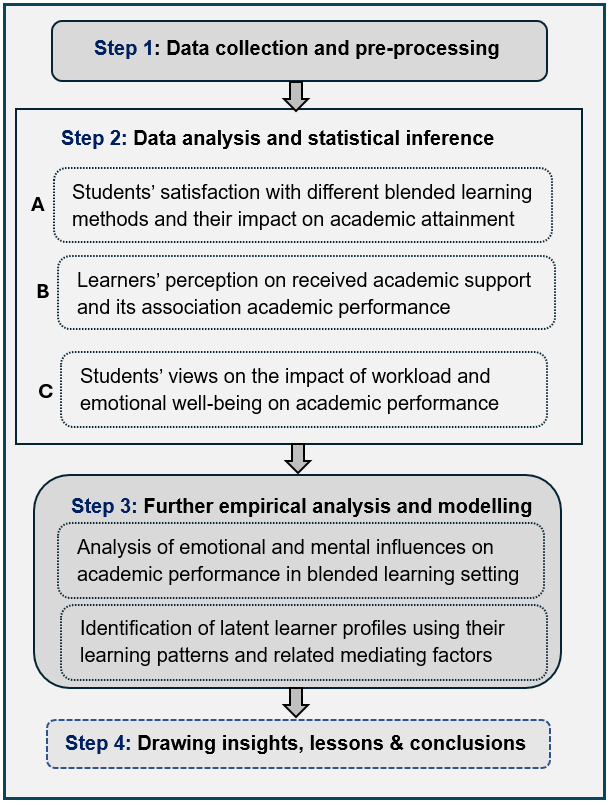}
 \caption{Research methodology workflow}
    \label{methodology}
 \vspace{-5pt}
\end{figure}

In this study, we have employed a four-step research implementation method as illustrated in \autoref{methodology}. The empirical data is first collected via a tailored questionnaire and pre-processed in step 1. Next, we have performed a set of data analysis and inference tasks to investigate the research questions/objectives (Step 2). Further consolidating examination has been conducted in step 3 with the aim of evidencing the joint effects of student motivation, workload, and mental state on academic achievement. Finally, the data analysis and statistical inference results were analysed and interpreted under the CoI framework to drawn practical insights and lessons (Step 4). The following sections detail the aforementioned research implementation steps. 

\subsection{Data Collection and preprocessing}
The key objective of the study was to understand lived experiences of full-time postgraduate students in UK universities following the adoption of BL mode, largley inherited from COVID-19 adaptations in Higher Education (HE). To capture these perspectives, this research used primary data collection through an online survey, which was considered an effective approach for the target participants as the postgraduate students are scattered across various universities in the UK. Through the survey, the study captured various aspects of student's lived experience with blended learning including satisfaction with different learning methods, agreement with academic support, percieved stress/study load changes, and the broader impact of all of these on their academic experiences. Table~\ref{DataDiscreption} shows description of the key data variables collected for the study and used in our analysis.

\begin{table}[t!]
 \centering
 \caption{Key collected variables used in the study's analysis and their data types}
 \label{DataDiscreption}
 \begin{tabular}{l@{\hskip 0.2in} l@{\hskip 0.2in} l@{\hskip 0.5in}}
        \toprule
\textbf{No} & \textbf{Data variable} & \textbf{Data type}\\
         \hline\addlinespace[2pt]
1 & Age in ranges 	& Ordinal (e.g., 18-24, etc) \\
2	& Gender &	Nominal (e.g. female, etc) \\
3	& Average study time in hours per day &	Numeric (e.g., 3 hrs, etc) \\
4	& Satisfaction with online live classes &	Ordinal/Likert scale (e.g. 1-5) \\
5	& Satisfaction with recorded lectures &	Ordinal/Likert scale (e.g. 1-5) \\
6	& Satisfaction with lecture materials 	& Ordinal/Likert scale (e.g. 1-5) \\
7	& Agreement with feedback on assignments &	Ordinal/Likert scale (e.g. 1-5) \\
8	& Agreement with feedback on questions &	Ordinal/Likert scale (e.g. 1-5) \\
9	& Agreement with feedback on exams	& Ordinal/Likert scale (e.g. 1-5) \\
10	& Perceived change/increase in stress &	Ordinal/Likert scale (e.g. 1-5) \\
11	& Perceived change/increase in study load &	Ordinal/Likert scale (e.g. 1-5) \\
12	& Perceived impact on learner psychology & Ordinal/Likert scale (e.g. 1-5) \\
13	& Improved student motivation for learning & Ordinal/Likert scale (e.g. 1-5) \\
14	& Self-assessed academic performance &	Ordinal/Likert scale (e.g. 1-3) \\
\vspace{-4pt}
 		\bottomrule
 	\end{tabular}
 \vspace{-10pt}
 \end{table}
 
\subsection{Data Analysis and Inference}\label{DataAnalysis}
 One of the key objectives of this study is to understand how students' academic performance (target) is affected by various factors including teaching and learning methods, study workload, learners' mental well-being and the support provided by teaching staff. This section presents an analysis performed on the survey data to investigate these objectives. First, we examine how demographic and behavioral factors affect the data composition and students' academic performance in blended learning contexts.

\subsubsection{Profile of the study participants}\label{Profiles}
\begin{figure}[h!]
\centering
\includegraphics[width=\columnwidth]{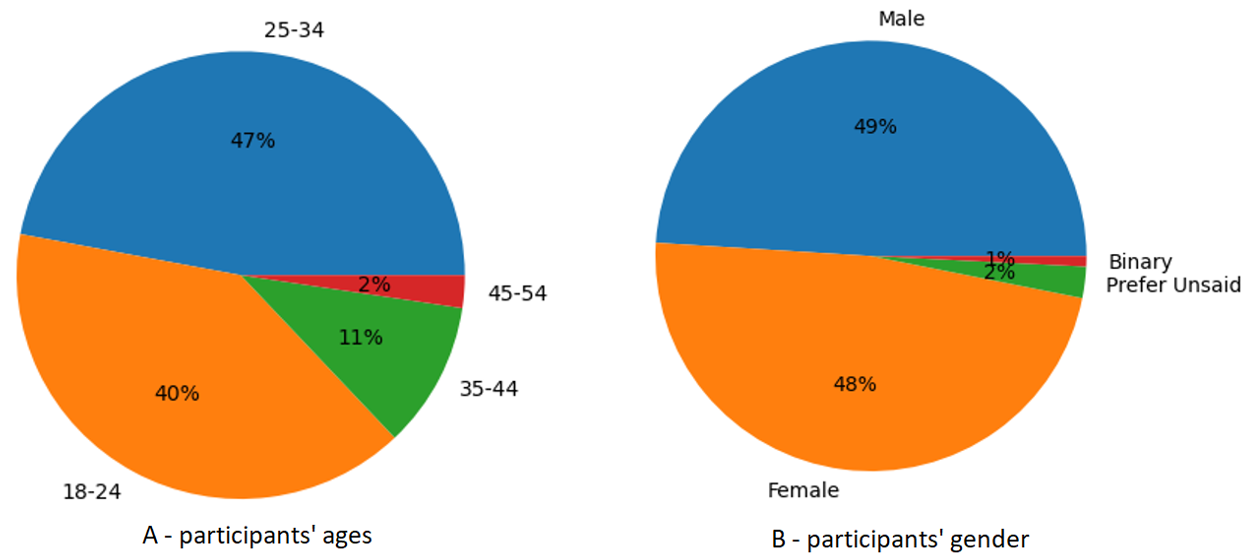}
 \caption{Age (A) and gender (B) distributions of the survey participants.}
    \label{participant_profile}
\end{figure}

Participation of the survey reflected a typical M.Sc students’ profile with ages ranging from under 24 to over 50. As illustrated in \autoref{participant_profile}, the 25–34-year-olds represented the largest age group (47\%), followed by 18–24-year-olds (40\%). Gender-wise, male, and female participant proportions were almost similar (m-49\%, f-48\%) with very few students either not stating their gender or identifying themselves as binary (1\%). 

\begin{figure}
\centering
\includegraphics[width=0.9\columnwidth]{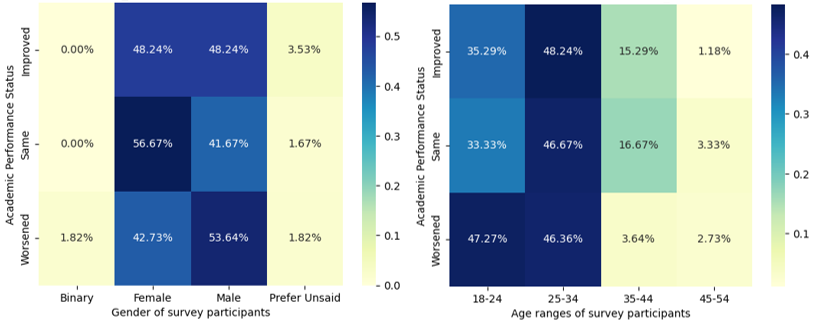}
 \caption{Distribution of study participants in terms of academic performance by gender and age group}
    \label{age-gender_vs_acper}
 \vspace{-8pt}
\end{figure}

Next, we have explored how the respondents' age and gender is related to academic performance (study’s target variable). In this and any subsequent analysis, we would like to note that the reported academic status categories are judgments by the survey participants as opposed to information taken from academic assessments, such as exams. \autoref{age_vs_acper} shows how academic performance of the different groups is affected. The results suggest that the academic performance of young students (18 - 34) were more negatively affected compared to their relatively older peers (35-54). This could mean that more experienced students who could have been in employment prior to joining the postgraduate program were more equipped with the knowledge, and skills required to mitigate the new ways of learning, e.g. fully online synchronous and asynchronous settings. In addition, and as illustrated in \autoref{gender_vs_acper} (cf. \autoref{age-gender_vs_acper}), male (53.64\%) students seem to be more slightly negatively affected by the blended learning adaptations compared to their female (42.73\%) cohort. 


\begin{table}[t!]
 \centering
 \caption{Relationship between age group and academic performance (total survey participants (n): 255)}
 \label{age_vs_acper}
 \begin{tabular}{l@{\hskip 0.5in} l@{\hskip 0.5in} l@{\hskip 0.5in} l@{\hskip 0.5in} l@{\hskip 0.5in}}
        \toprule
        \textbf{Age groups} & \multicolumn{3}{c}{\textbf{Academic performance}} \\
        \cmidrule{2-4}
         &  Improved&  Unchanged & Worsened\\
         \hline
         18 - 24 & 30 & 20 & 52\\
         25 - 34	& 41 &	28	& 51\\
         35 - 44 &	13 &	10	& 4\\
         45 - 54	& 1 &	2	& 3\\
         Total &	85 &	60 	& 110 \\
 		\bottomrule
 	\end{tabular}
 \vspace{-3pt}
 \end{table}

\begin{table}
 \centering
 \caption{Association between gender and academic performance (total survey participants (n): 255)}
 \label{gender_vs_acper}
 \begin{tabular}{l@{\hskip 0.5in} l@{\hskip 0.5in} l@{\hskip 0.5in} l@{\hskip 0.5in} l@{\hskip 0.5in}}
        \toprule
        \textbf{Gender} & \multicolumn{3}{c}{\textbf{Academic performance}} \\
        \cmidrule{2-4}
         &  Improved&  Unchanged & Worsened\\
         \hline
Female &	41	& 34 &	47 \\
Male	& 41 &	25 &	59 \\
Unspecified	& 3	& 1	& 2 \\
Binary &	0 &	0&	2 \\
Total &	85 &	60 	& 110 \\
 		\bottomrule
 	\end{tabular}
 \vspace{-10pt}
 \end{table}

\begin{figure}
\centering
\includegraphics[width=0.9\columnwidth]{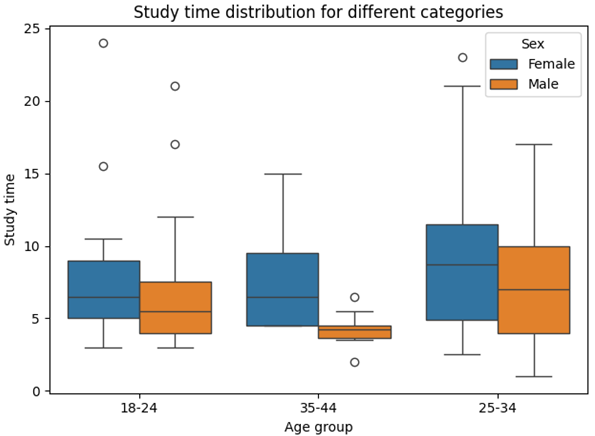}
 \caption{Comparison of the study times for different student groups based on the sex and age group variables: rare classes ($n<10$) are excluded for lack statistical interpretability}
    \label{studytime}
 \vspace{-8pt}
\end{figure}

A relevant attribute that our research has captured is the students’ study time in terms of  the average hours they spend in learning each day. Using this variable, we have investigated the existence of any differences in study patterns between various student groups. For example, \autoref{studytime} shows that female students spend relatively more time in their studies compared to male peers, which may explain their relatively better academic performance compared to male students (\autoref{gender_vs_acper}). The chart also demonstrates that the average learning time of the  ‘25-34’ year-old student group is higher compared to the rest of their mates.  
To confirm whether the results -- in terms of the study time differences between different learner groups -- are statistically significant, we run a t-test comparing the corresponding data samples. As a result, we found that the average study hours by male students is significantly lower than similar time by the fellow female learners ($p<0.01$, t-test). The test has also found similar conclusion for the learning hours of the different age groups, i.e., the existence of a statistically significant difference in mean study times across the three age groups ($p<0.05$, t-test). 

\subsubsection{To what extent is academic achievement in BL dependent on satisfaction?} \label{Sat_with_LT-methods1} 
One of the key questions that our study investigated is to confirm the existence of any dependence between students' level of satisfaction with different learning activities and their experience relating to the academic performance. Relating to this, our survey has asked students to assess their level of satisfaction with the different BL methods. The analysis based on the responses (n=255) of this question has shown that - of the three used T \& L strategies - learners were mostly contented with online live classes followed by pre-recorded lectures with approximately 55\% and 47\% respectively stating that they are (highly) satisfied with them as shown in \autoref{sat_with_LT-methods}. The table summarizes the satisfaction ratings expressed by surveyed students and the proportions that provided each rating. 

\begin{table}
 \centering
 \caption{Distributions of student satisfaction with different blended learning methods and resources: online live classes, pre-recorded lectures, and lecture materials; 'H' in the column headers represents 'Highly'}
 \label{sat_with_LT-methods}
 \begin{tabular}{l@{\hskip 0.1in} c@{\hskip 0.1in} c@{\hskip 0.1in} c@{\hskip 0.1in} c@{\hskip 0.1in} c@{\hskip 0.1in}}
        \toprule
        \textbf{L \& T Methods} & \multicolumn{5}{c}{\textbf{Students' ratings and the number who gave each rating}} \\
        \cmidrule{2-6}
         &  H. Dissatisfied&  Dissatisfied & Neutral &  Satisfied & H. Satisfied\\
         \hline
Online live classes	& 10 &	51 &	55 &	122 &	17 \\
Pre-recorded lect. &	14	& 65	& 56	& 100	& 20 \\
Lecture materials &	17	& 95	& 70	& 62	& 11 \\
\vspace{-0.05in}
 		\bottomrule
 	\end{tabular}
 \vspace{-5pt}
 \end{table}

The noticeable relative higher popularity of live sessions could be attributed to the fact that it provides real-time interaction with educators where students can receive instant support and feedback on queries. Similarly, pre-recorded lectures are largely designed as brief recordings with their shorter duration perhaps motivating students to engage.  The least contented item according to the survey study is the lecture materials (e.g. slides, notes, etc) which usually requires students to study larger materials asynchronously.  It is not surprising for students studying an extensive 12-month M.Sc course taking multiple modules each term, some of which may be in part-time employment, to engage less in such time consuming learning resources.  
\begin{figure}
\centering
\includegraphics[width=\columnwidth]{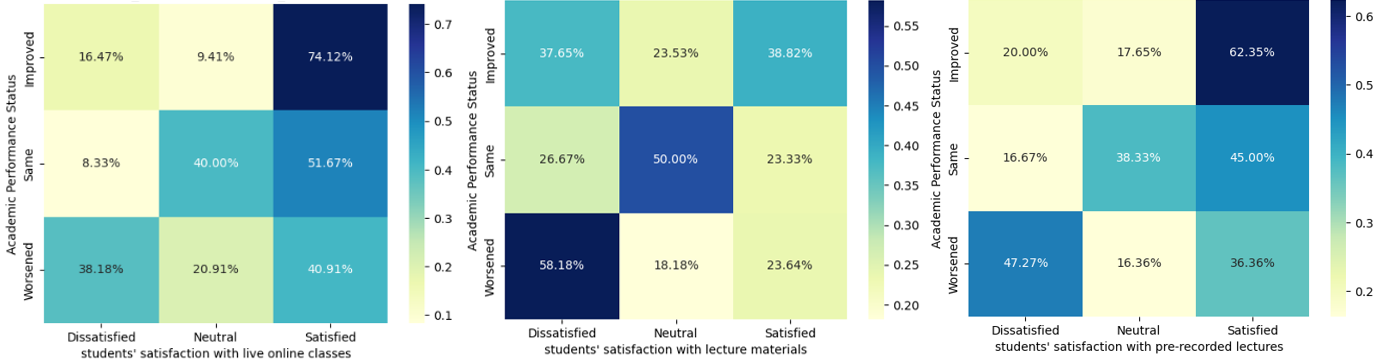}
 \caption{Student satisfaction with different blended learning methods: academic performance percentages for online live classes, pre-recorded lectures, and lecture materials}
    \label{fig:sat_with_LT-methods}
 \vspace{-8pt}
\end{figure}

\autoref{fig:sat_with_LT-methods} illustrates the breakdown of the satisfaction levels for different levels of academic performance. It indicates some disparities, suggesting that the larger percentage of the surveyed students whose performance \emph{improved} (62\%) were (highly) satisfied with \emph{online live classes} (74.12\%) and \emph{pre-recorded lectures} (62.23\%). It also shows that the higher proportion of students whose performance \emph{worsened} were unhappy with the three learning methods, with the lecture materials being the least satisfied (58.18\%). 

\subsubsection{How does teaching staff support affects students’ academic performance?} \label{StaffSupport}
Our study has also explored how students feel about the academic support that the university lecturers provided during their studies. Specifically, students have been asked to rate their views and agreement with three academic support aspects by the lecturers: feedback on assignments, answering their questions, and guidance on key assessments, e.g. exams. \autoref{tab:agree_with_sup-types} summarizes the 5-level distribution of student agreement ratings on the three learning support activities. 

\begin{table}
 \centering
 \caption{Breakdown of student agreement with feedback on different academic support activities by teaching staff; the notation 'S' in the column headers represents 'Strongly'}
 \label{tab:agree_with_sup-types}
 \begin{tabular}{l@{\hskip 0.1in} c@{\hskip 0.1in} c@{\hskip 0.1in} c@{\hskip 0.1in} c@{\hskip 0.1in} c@{\hskip 0.1in}}
        \toprule
        \textbf{L \& T support type} & \multicolumn{5}{c}{\textbf{Students' ratings and the number who gave each rating}} \\
        \cmidrule{2-6}
         &  S. Disagree&  Disagree & Neutral &  Agree & S. Agree\\
         \hline
Assignment feedback	& 10	& 40	& 54	& 138	& 13 \\
Answering questions	& 5	& 79	& 63	& 95	& 13 \\
Exam advise/feedback &	5	& 35	& 42	& 153	& 20 \\
\vspace{-0.05in}
 		\bottomrule
 	\end{tabular}
 \vspace{-5pt}
 \end{table}
 
\begin{figure}
\centering
\includegraphics[width=\columnwidth]{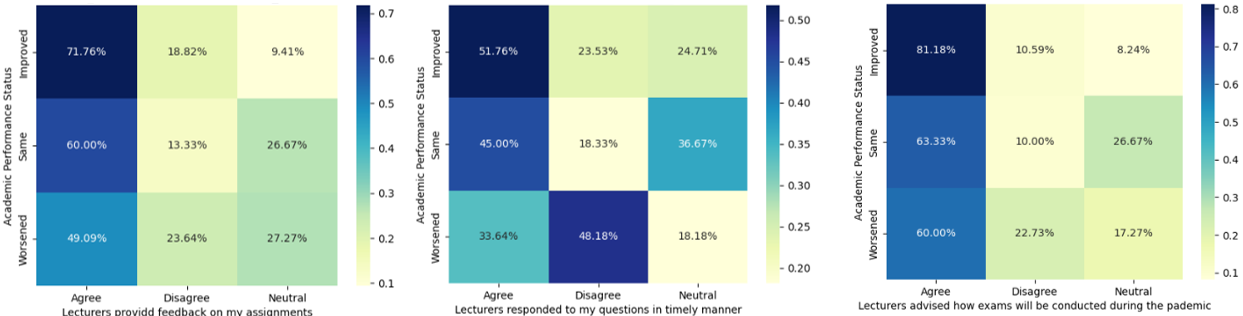}
 \caption{Student views on different academic support tasks provided by the university lecturers: percentage agreement with feedback on assignments, questions and exams}
    \label{fig:agree_with_sup-types}
 \vspace{-8pt}
\end{figure}

Notably, 60\% of the surveyed students agree that the lectures provided feedback on their assignments and clear guidance on their exams. However, that sentiment drops for the third question `Lecturers responded to my questions in a timely manner' to about 42\% with approximately 33\% of the participating students disagreeing with it (cf. 19.6\% for feedback on assignments and 15.6\% for exam guidance). Crisis (e.g., COVID-19) induced adaptations in HE institutions overwhelm academic staff requiring them to, among other things, simultaneously learn new technologies and adjust their teaching materials and activities for blended learning. This increased workload could have affected the available time, leading to delays in responding to student queries which could have resulted in such a learner feeling.

In addition and based on the further breakdown of student agreement levels for different academic performance groups in \autoref{fig:agree_with_sup-types}, there is a clear dependence between the different academic support activities and the higher proportions of students who agree the effectiveness of these learning support activities by teaching staff in blended learning contexts. This is more evident from the number of students who voted that their academic performance has either \emph{mproved} or \emph{unchanged}.

\subsubsection{What influence do study workload and stress have on academic achievement?}\label{wkloadStress}
\begin{figure}
\centering
\includegraphics[width=0.9\columnwidth]{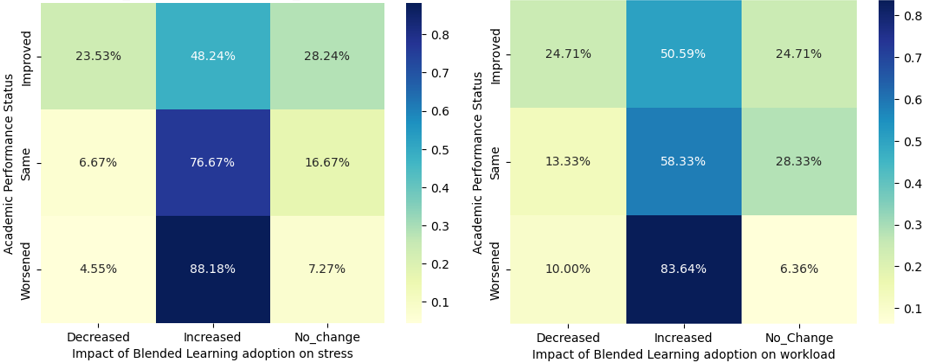}
 \caption{Changes in study workload and stress \& their perceived impact on students' academic performance}
    \label{fig:workload-stress}
 \vspace{-8pt}
\end{figure}

Two other student experience factors that our study investigated are stress and workload, and their impact on academic achievement. Related to this, \autoref{tab:wkload-stress} provides the 5-level sentiment distributions of these two factors. It shows that over 72\% and 66\% of the total surveyed students felt that the introduced blended learning adaptations either slightly (47\%, 53\%) or significantly (25\%, 14\%) increased their stress levels and workload respectively. Besides, we further inspected whether changes in stress levels and workload, on the one hand, were associated with academic achievement levels, on the other. From the visual insights illustrated in \autoref{fig:workload-stress}, it is very clear that increased stress and workload is linked to low academic outcome i.e., over 80\% of the students who declared that they academic performance worsened also stated their workload and stress levels increased. 
\begin{table}
 \centering
 \caption{Changes in study workload and stress among the surveyed study participants; the notations 'Si' and 'Sl' in the column headers respectively represent 'Significantly' and 'Slightly'}
 \label{tab:wkload-stress}
 \begin{tabular}{l@{\hskip 0.05in} c@{\hskip 0.05in} c@{\hskip 0.05in} c@{\hskip 0.05in} c@{\hskip 0.05in} c@{\hskip 0.05in}}
        \toprule
        \textbf{\makecell{Workload \\ and Stress}} & \multicolumn{5}{c}{\textbf{Students' ratings and the number who gave each rating}} \\
        \cmidrule{2-6}
         & Si. decreased & Sl. decreased & No change &  Sl. increased & Si. increased\\
         \hline
Stress &	5 &	24 &	42	& 120 &	64 \\
Workload &	3 &	37	& 45 &	135	& 35 \\
\vspace{-0.05in}
 		\bottomrule
 	\end{tabular}
 \vspace{-5pt}
 \end{table}

\subsubsection{Testing the dependence of student experience factors on academic performance}\label{HypoTesting}
In addition to the descriptive analysis and results presented in Sections~\ref{Profiles}-\ref{wkloadStress}, we have performed several statistical tests between the study's key target variable (academic performance) and the explanatory variables that captured the student experience. Especially, we used Likert scale ratings ~\citep{sullivan2013analyzing} to collect student experience data in blended learning context. Since the study data are predominantly ordinal, we have selected several nonparametric test methods (e.g., spearman correlation and Mann-Whitney) to conduct the analysis. 

\paragraph{Analysis based on Spearman test}
For our first statistical test, we have performed Spearman correlation (\autoref{SpearmanEq}) test to assess the monotonic relationships between the studied L\&T factors and the academic performance as summarized in Table~\ref{SpearmanTest}. Clearly, all the test findings obtained indicate significant results. In other words, the association between all investigated student experience factors – satisfaction with different teaching and learning activities, agreement ratings with different academic support tasks by teaching staff, and changes in stress and workload – and academic performance levels were found to be statistically significant. The results suggest that satisfaction with synchronous live classes and changes in workload and stress are the most significant determinants for academic attainment, which is consistent with the explanatory insights presented in Sections~\ref{Profiles}-\ref{wkloadStress}. Notably, the study workload and stress indicators are negatively associated with the academic performance highlighting that their increase undermines academic achievement. The low coefficient scores (middle column) do not imply that the associations are non-significant but that correlation values are constrained by the limited ranges of the used ordinal/Likert scale data (e.g. 1-3, 1-5). This is also confirmed through the corresponding p values which all fall below the significant level (e.g. 0.05).  

\begin{table}[t!]
 \centering
 \caption{Spearman test results on the association between academic performance and student experience factors: all scores are rounded to 3 SF}
 \label{SpearmanTest}
 \begin{tabular}{l@{\hskip 0.2in} l@{\hskip 0.2in} l@{\hskip 0.2in} l@{\hskip 0.2in}}
        \toprule
        \textbf{Compared factors/variables} & \multicolumn{3}{c}{\textbf{Spearman test results}}\\
        \cmidrule{2-4}
         &  Coefficient &  P value\\
         \hline\addlinespace[2pt]
Recorded lectures vs Academic performance &	0.258	& \num{3.054e-05} \\
Live online classes vs Academic performance & 0.312	& \num{3.829e-07} \\
Lecture Materials vs Academic performance &	0.200	& \num{1.294e-03} \\
Answering questions vs Academic performance &	0.225 &	\num{2.766e-04} \\
Assignment feedback vs Academic performance	& 0.183	& \num{3.405e-03} \\
Exam feedback vs Academic performance &	0.226	& \num{2.707e-04} \\
Stress level vs Academic performance &	-0.382	& \num{2.6705e-10} \\
Study workload vs Academic performance &	-0.281	& \num{5.011e-06} \\

\vspace{-4pt}
 		\bottomrule
 	\end{tabular}
 \vspace{-10pt}
 \end{table}

\vspace{-5pt}
\begin{equation}\label{SpearmanEq}
\rho_s = 1 - \frac{6 \sum d_i^2}{n(n^2 - 1)}
\vspace{-5pt}
\end{equation}
\paragraph{Analysis based on Mann-Whitney U test}
In addition to exploring the association between student academic performance and learning determinants (Table \ref{SpearmanTest}), we have also investigated whether the average scores of students of different levels of academic performance are statistically different from each other in relation to the various aspects of student experience. For example, are students with \emph{improved} academic performance more satisfied with blended learning techniques compared to those whose performance \emph{worsened}? Since the study's analytic data are largely ordinal type and the corresponding numeric ratings cannot be treated as continuous data, we have opted for Mann-Whitney U (MWU) test. We have selected the median as comparison statistic on which the statistical test is based because interpreting median agreement or satisfaction would be more meaningful for the likert scale ordinal data compared to their mean value~\citep{sullivan2013analyzing}. In addition, some studies have found that the Mann-Whitney test is more powerful for Likert scale data compared to parametric tests, particularly with smaller sample sizes and unequal group sizes~\citep{csimcsek2023power}.
\vspace{-10pt}
\begin{equation}
U = \min(U_1, U_2); \,
U_1 = n_1 n_2 + \frac{n_1(n_1 + 1)}{2} - \textstyle \sum_{}^{}R_1; \,
U_2 = n_1 n_2 + \frac{n_2(n_2 + 1)}{2} - \sum_{}^{}R_2
\vspace{-5pt}
\end{equation}
Where \( n_1 \), \( n_2 \) are the sample sizes of the two groups, and \( \textstyle \sum_{}^{}R_1\), \( \sum_{}^{}R_2\) are the sums of median ratings for each group.


 \begin{table}[h!]
 \begin{flushleft}
 \caption{Mann Whitney U (MWU) test results on the difference of medians (satisfaction) between different academic performance groups for the online live classes: all scores are rounded to 3 SF}
 \label{tab:WMU-live_class}
 \begin{tabular}{l@{\hskip 0.05in} l@{\hskip 0.05in} l@{\hskip 0.05in} l@{\hskip 0.05in}}
        \toprule
        \textbf{Paired AP levels} & \textbf{Paired Rating Medians} & \multicolumn{2}{c}{\textbf{MWU test results}} \\
        \cmidrule{3-4}
         &  &  Statistic & P value\\
         \hline       
Worsened(W)-Same(S)	& W=3(neutral); S=4(satisfied) &	2485.0 &	\num{4.921e-03}\\
Worsened(W)-Improved(I)	& W=3(neutral); I=4(satisfied)	& 2982.0 &	\num{3.601e-06}\\
Same(S)-Improved(I)	& S=4(satisfied); I=4(satisfied)	& 2012.5	& \num{1.724e-02}\\
 		\bottomrule
 	\end{tabular}
     \end{flushleft}
 \vspace{-10pt}
 \end{table}

 \begin{table}
 \begin{flushleft}
 \caption{Mann Whitney U test results on the difference of medians (satisfaction) between different academic performance groups for pre-recorded lectures: all scores are rounded to 3 SF}
 \label{tab:WMU-RecordedLecture}
 \begin{tabular}{l@{\hskip 0.05in} l@{\hskip 0.05in} l@{\hskip 0.05in} l@{\hskip 0.05in}}
        \toprule
        \textbf{Paired AP levels} & \textbf{Paired Rating Medians} & \multicolumn{2}{c}{\textbf{MWU test results}} \\
        \cmidrule{3-4}
         &  &  Statistic & P value\\
         \hline       
Worsened(W)–Same(S)	& W=3(neutral); S=3(neutral)&	2547.5&	\num{1.039e-02}\\
Worsened(W)–Improved(I)	& W=3(neutral); I=4(satisfied)&	3250.5&	\num{1.376e-04}\\
Same(S)–Improved(I)	& S=3(neutral); I=4(satisfied)	& 2149.0	& \num{8.646e-02}\\
 	\bottomrule
 	\end{tabular}
     \end{flushleft}
 \vspace{-10pt}
 \end{table}

 \begin{table}[h!]
 \begin{flushleft}
 \caption{Mann Whitney U test results on the difference of medians (satisfaction) between different academic performance groups for Learning materials: all scores are rounded to 3 SF}
 \label{tab:WMU-LectureMaterials}
 \begin{tabular}{l@{\hskip 0.05in} l@{\hskip 0.05in} l@{\hskip 0.05in} l@{\hskip 0.05in}}
        \toprule
        \textbf{Paired AP levels} & \textbf{Paired Rating Medians} & \multicolumn{2}{c}{\textbf{MWU test results}} \\
        \cmidrule{3-4}
         &  &  Statistic & P value\\
         \hline       
Worsened(W)–Same(S)	& W=2(Dissatisfied); S=3(neutral)&	2620.0&	\num{1.963e-02}\\
Worsened(W)–Improved(I)	& W=2(Dissatisfied); I=3(neutral)&3561.0& \num{2.613e-03}\\
Same(S)–Improved(I)	& S=3(neutral); I=3(neutral)& 2364.0&	0.438\\
 	\bottomrule
 	\end{tabular}
     \end{flushleft}
 \vspace{-12pt}
 \end{table}

From the test findings summarized in Tables \ref{tab:WMU-live_class}-\ref{tab:WMU-LectureMaterials}, we note that all test results are statistically significant ($p<0.05$), except those of the \emph{Same} and \emph{Improved} groups for the \emph{pre-recorded} and \emph{lecture material} resources ($p > 0.05$). In other words, the median satisfaction rating of the students whose academic performance \emph{did not change} or \emph{improved} is statistically significantly different (i.e, higher) from those with \emph{worsened} performance. The notations AP and MWU  in Tables \ref{tab:WMU-live_class}-\ref{tab:WMU-LectureMaterials} represent \emph{academic performance} and \emph{Mann-Whitney U} test respectively. The results also show that the students were least satisfied with \emph{lecture materials}, and the median difference is intuitively more significant between the \emph{Worsened} and \emph{Improved} groups. Overall, the outcomes of the Mann-Whitney test appear to be clearly consistent with the previous results including the Spearman test findings. 

 \begin{table}[h!]
 \begin{flushleft}
 \caption{Statistical test results (MWU) of the difference of medians (agreement) between different academic performance groups relating to tutor feedback on student assignments: all scores are rounded to 3 SF}
 \label{tab:AssignmentFeedback}
 \begin{tabular}{l@{\hskip 0.05in} l@{\hskip 0.05in} l@{\hskip 0.05in} l@{\hskip 0.05in}}
        \toprule
        \textbf{Paired AP levels} & \textbf{Paired Rating Medians} & \multicolumn{2}{c}{\textbf{MWU test results}} \\
        \cmidrule{3-4}
         &  &  Statistic & P value\\
         \hline       
Worsened(W)–Same(S)&	W=3(neutral); S=4(Agree)& 2853.0	& \num{1.142e-01}\\
Worsened(W)–Improved(I)	& W=3(neutral); I=4(Agree)	& 3680.5	& \num{5.399e-03}\\
Same(S)–Improved (I) &	S=4(Agree); I=4(Agree)	& 2279.0	& \num{2.181e-01}\\
 	\bottomrule
 	\end{tabular}
     \end{flushleft}
 \vspace{-10pt}
 \end{table}

  \begin{table}[h!]
 \begin{flushleft}
 \caption{Statistical test results (MWU) of the difference of medians (agreement) between different academic performance groups relating to tutor feedback on exams: all scores are rounded to 3 SF}
 \label{tab:ExamFeedback}
 \begin{tabular}{l@{\hskip 0.05in} l@{\hskip 0.05in} l@{\hskip 0.05in} l@{\hskip 0.05in}}
        \toprule
        \textbf{Paired AP levels} & \textbf{Paired Rating Medians} & \multicolumn{2}{c}{\textbf{MWU test results}} \\
        \cmidrule{3-4}
         &  &  Statistic & P value\\
         \hline       
Worsened(W)–Same(S)	& W=4(Agree); S=4(Agree)	&2912.5	& \num{1.614e-01}\\
Worsened(W)–Improved(I)	& W=4(Agree); I=4(Agree)	&3439.5	& \num{2.750e-04}\\
Same(S)–Improved(I)	& S=4(Agree); I=4(Agree)	&2145.5	& \num{6.192 e-01}\\
 		\bottomrule
 	\end{tabular}
     \end{flushleft}
 \vspace{-10pt}
 \end{table}

  \begin{table}[h!]
 \begin{flushleft}
 \caption{Statistical test results (MWU) of the difference of medians (agreement) between different academic performance groups relating to tutor answers to student questions: all scores are rounded to 3 SF}
 \label{tab:QA-Feedback}
 \begin{tabular}{l@{\hskip 0.05in} l@{\hskip 0.05in} l@{\hskip 0.05in} l@{\hskip 0.05in}}
        \toprule
        \textbf{Paired AP levels} & \textbf{Paired Rating Medians} & \multicolumn{2}{c}{\textbf{MWU test results}} \\
        \cmidrule{3-4}
         &  &  Statistic & P value\\
         \hline       
Worsened(W)–Same(S)&	W=3(neutral); S=3(neutral)	& 2521.5&	\num{7.320e-03}\\
Worsened(W)–Improved(I)	& W=3(neutral); I=4(Agree) &	3425.5&	\num{7.737e-04}\\
Same(S)–Improved(I)	& S=3(neutral); I=4(Agree)&	2361.5&	\num{4.23e-01}\\
 	\bottomrule
 	\end{tabular}
     \end{flushleft}
\vspace{-10pt}
 \end{table}

Similarly, Tables \ref{tab:AssignmentFeedback}-\ref{tab:QA-Feedback} summarize statistical test results of the difference between median agreements ratings by different student performance groups on three key tutor support types: feedback on assignments, exams, and student queries. In Tables \ref{tab:AssignmentFeedback} and \ref{tab:ExamFeedback}, we can see that the median difference test results are not significant – although student agreements on staff responses to their queries differ  –  except that of the difference between \emph{Worsened} and \emph{Improved} groups. It is also clear that the impact on students' academic performance is related to their received support in terms of answering their questions (Table~\ref{tab:QA-Feedback}). One of the key conclusions that can be drawn from the results is that all three types of feedback and support from the educators underpin student academic performance, which suggests that academic improvements depend on the level of academic support provided.
  \begin{table}[h!]
 \begin{flushleft}
 \caption{Statistical test results (MWU) of the difference of medians between different academic performance groups relating to study workload impact due to adopted Blended learning: all scores are rounded to 3 SF}
 \label{tab:wkload-vs-ap}
 \begin{tabular}{l@{\hskip 0.05in} l@{\hskip 0.05in} l@{\hskip 0.05in} l@{\hskip 0.05in}}
        \toprule
        \textbf{Paired AP levels} & \textbf{Paired Rating Medians} & \multicolumn{2}{c}{\textbf{MWU test results}} \\
        \cmidrule{3-4}
         &  &  Statistic & P value\\
         \hline       
Worsened(W)–Same(S)& W=4(Slightly\_I); S=4(Slightly\_I) &	4034.0 & \num{7.062 e-03}\\
Worsened(W)–Improved(I)& W=4(Slightly\_I); I=4(Slightly\_I) & 6235.0 & \num{1.065e-05}\\
Same(S)–Improved(I)	& S=4(Slightly\_I); I=4(Slightly\_I)&	2886.0 &	\num{1.550e-01}\\
 	\bottomrule
 	\end{tabular}
     \end{flushleft}
 \vspace{-10pt}
 \end{table}

   \begin{table}[h!]
 \begin{flushleft}
 \caption{Statistical test results (MWU) of the difference of medians between different academic performance groups relating to stress impact due to the adopted Blended learning: all scores are rounded to 3 SF}
 \label{tab:stress-vs-ap}
 \begin{tabular}{l@{\hskip 0.05in} l@{\hskip 0.05in} l@{\hskip 0.05in} l@{\hskip 0.05in}}
        \toprule
        \textbf{Paired AP levels} & \textbf{Paired Rating Medians} & \multicolumn{2}{c}{\textbf{MWU test results}} \\
        \cmidrule{3-4}
         &  &  Statistic & P value\\
         \hline       
Worsened(W)–Same(S) &	W=4(Slightly\_I); S=4(Slightly\_I) &	4030.5	& \num{8.576e-03}\\
Worsened(W)–Improved(I) &	W=4(Slightly\_I); I=3(No change) &	6854.5	& \num{3.385e-09}\\
Same(S)–Improved(I)	& S=4(Slightly\_I); I=3(No change) &	3314.0 &	\num{1.189e-03}\\
 	\bottomrule
 	\end{tabular}
     \end{flushleft}
     \vspace{-8pt}
 \end{table}
 
Finally, the results in Tables~\ref{tab:wkload-vs-ap} and \ref{tab:stress-vs-ap} confirm the initial high-level findings in Table~\ref{tab:wkload-stress} (cf. Table~\ref{SpearmanTest}) where we found statistical association between students' self-evaluated academic performance and BL induced workload and stress. The test outcomes evidence that the level of BL impact on both stress and study workload is statistically significantly different from one student group to another. However, no statistical differences were found in the experience of the study workload for students whose academic performance did \emph{not change} against those with \emph{improved} achievement. Overall, the findings suggest that increased stress and workload both negatively affect learners' academic attainments.

\paragraph{Analysis based on Ordinal Regression}
In addition to the previously presented exploratory and confirmatory analysis, we applied ordinal regression based on the most correlated predicting factors and academic performance as the target (\autoref{Logistic Regression Results}). The rationale here is to both further understand the predictive interaction between social and cognitive presences with learning outcomes, and identify the parsimonious model and variables that can explain academic performance in a blended learning setting. Student motivation and future goal variables can be seen as both prerequisites for and outcomes of strong cognitive presence. Motivated students with clear goals are more likely to engage in effective learning and reflective inquiry.
\begin{table}[h!]
 \centering
 \caption{Parsimonious model results using ordinal logistic regression based on most significant variables}
 \label{Logistic Regression Results}
 \begin{tabular}{l@{\hskip 0.4in} c@{\hskip 0.4in} c@{\hskip 0.4in} c@{\hskip 0.4in} c@{\hskip 0.3in} c@{\hskip 0.3in}}
        \toprule
        \textbf{Variable name}  \\
        \cmidrule{2-5}
         &  Coeff &  Std err & Z &   P value \\
         \hline
Live online classes	& 0.3495 &	0.083 &	4.222 &	0.000 \\
Study workload &	-0.2280	& 0.092	& -2.479	& 0.013	 \\
Stress level &	-0.2914	& 0.100	& -2.918	& 0.004	 \\
Future goals &	0.1949	& 0.098	& 1.983	& 0.047	 \\
Student motivation &	0.1914	& 0.098	& 2.815	& 0.005	 \\
Psychological effect &	-0.1880	& 0.099	& -1.908	& 0.056	 \\
\vspace{-0.05in}
 		\bottomrule
 	\end{tabular}
\vspace{-5pt}
 \end{table}
The results (\autoref{Logistic Regression Results}) show that higher engagement in live classes, clarity on future goals, and stronger student motivation are all associated with better academic performance. Higher perceived workload, greater stress levels, and negative psychological affect are all associated with poorer academic performance, which is consistent with the previous confirmatory results (\autoref{SpearmanTest}). In terms of the coefficient, live classes seems to have the most effect on academic performance in an online setting and high levels of stress affects learning success most negatively. 

\section{Discussion and CoI analysis of results}
This section presents an empirical consolidation discussion -- through moderation, mediation, and profile analysis based on parsimonious learning factors -- and reflections on the main research findings linked with the principles of the CoI framework. For this, Table~\ref{tab:CoI-mapping} lists the key study variables used in our analysis grouped by the corresponding CoI presences. The context variables (last row in Table~\ref{tab:CoI-mapping}) are not strictly part of the framework presences, but relevant moderating factors used to understand the relationships between the different framework dimensions (Section~\ref{Profiles}).  


 \begin{table}[h!]
 \caption{Mapping our research data variables to corresponding Community of Inquiry (CoI) presences}
 \label{tab:CoI-mapping}
\begin{tabular}{p{3.5cm}@{\hskip 0.05in} p{10cm}}
\toprule
\textbf{CoI Dimension} & \textbf{Relevant variables capturing that dimension} \\
\hline
Teaching Presence & 
Student satisfaction with learning methods (lecture materials, recordings, online live class),  Learner rating on lecturer feedback on assignments, learner questions, and exams \\
Social Presences & 
Student satisfaction with online live classes, perceived impact on learner psychology \\
Cognitive Presence & 
Self-assessed academic performance, perceived change/increase in workload and stress, improved student motivation for learning \\
Context variables & 
Learner age, gender, and average study time \\
\bottomrule
\end{tabular}
\vspace{-10pt}
 \end{table}


\subsection{Mediation and moderation effects of resilience factors}
\begin{figure}[h!]
\centering
\includegraphics[width=\columnwidth]{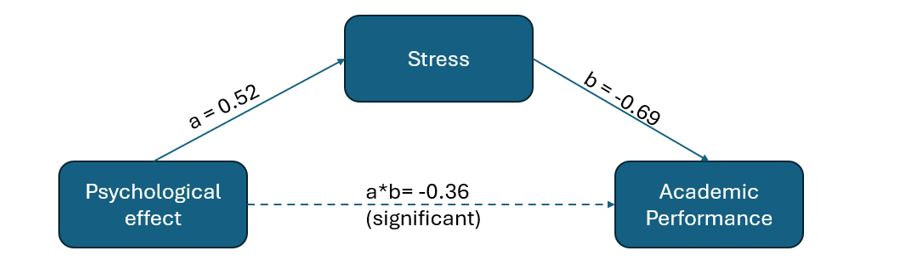}
 \caption{Impact of stress as a mediator between psychological effect and academic performance}
    \label{Mediation}
\vspace{-10pt}
\end{figure}

Our research has also investigated the mediation and moderation effects of wellbeing factors on academic performance, and we present here some of the most relevant quantitative relationships between the significant variables. \autoref{Mediation} demonstrates the mediation effect of stress on academic achievement. It suggests that the effect of psychological effect on stress levels (a = 0.52) is positive, showing that higher negative influence (e.g., anxiety, depressive symptoms) leads to elevated stress. In contrast, the effect of stress on academic performance (b = -0.69) is strongly negative, indicating that stress significantly undermines performance. The indirect effect (a × b = -0.36) is also negative and significant, confirming that stress is a key mechanism through which psychological distress lowers performance. This highlights stress as a critical mediator, reinforcing established findings that unmanaged stress is a primary channel linking mental health to academic difficulties ~\citep{harper2024face, aristovnik2020impacts}.

\begin{figure}[h!]
\centering
\includegraphics[width=\columnwidth]{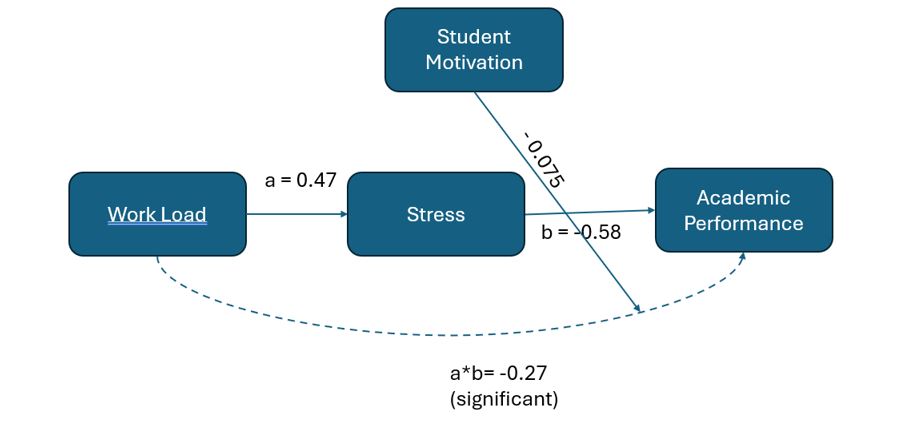}
 \caption{Impact of student motivation as a moderator between workload and academic performance mediated by stress}
    \label{Moderation}
\vspace{-10pt}
\end{figure}

Besides, the effect of workload (which can be considered a direct outcome of teaching presence) on stress (a = 0.47) is positive, meaning heavier workloads significantly increase stress (\autoref{Moderation}). Stress, in turn, negatively predicts academic performance (b = -0.58), confirming stress as a detrimental mediator. The indirect effect (a × b = -0.27) is substantial and negative, suggesting that workload harms performance primarily through heightened stress. The interaction term (-0.075) indicates that the relationship between workload and performance is moderated by student motivation: when motivation is higher, the negative effect of workload on performance via stress is slightly reduced. In general, high stress and negative psychological affect can be symptoms of low social presence feelings of isolation, which are common stressors in blended online learning~\citep{richardson2012psychological}.

The moderation and mediation finding (Workload→Stress→Performance, moderated by motivation) illustrated in \autoref{Moderation} is informative, as the instructor's course design (workload) can create stress that impairs performance and could lead to a breakdown of teaching presence. However, high student motivation which, as we've seen, can be boosted by the instructor acts as a protective buffer. A well-designed course should aim to make the workload challenging but manageable, thus preventing this negative pathway from activating in the first place. In addition, the path: Psychological distress→Stress→Performance (\autoref{Mediation}) demonstrates that a lack of well-being directly harms academic outcomes. Although CoI framework does not have an 'emotional presence', a strong social presence where students feel supported and connected can be the primary mechanism for mitigating the stress and isolation that may lead to poor outcomes.

Overall, a successful online blended learning experience can be said to rely on teaching presence (manageable workload, motivating feedback, etc) that fosters a supportive social presence. This environment serves as a crucial buffer against the stress and psychological issues that primarily undermine cognitive presence and, ultimately, academic performance.

\subsection{Latent Profile Analysis}
Although the interpretation of the obtained results through the CoI framework provides us with various relevant useful insights, we did some further analysis using Latent Profile Analysis (LPA) method to identify the heterogeneity of student experience in a blended learning setting. LPA is a statistical method which is used to classify subgroups within a population based on the patterns of their responses across a set of variables. It's a form of mixture modelling that assumes individuals can be grouped into distinct classes or profiles, each characterized by a unique configuration of variable means and variances~\citep{spurk2020}.  Based on the results of our analysis we found four significant student profiles (k = 4) and labelled them as profile 0 (baseline), 1, 2 and 3.

\begin{figure}[h!]
\centering
\includegraphics[width=0.9\columnwidth]{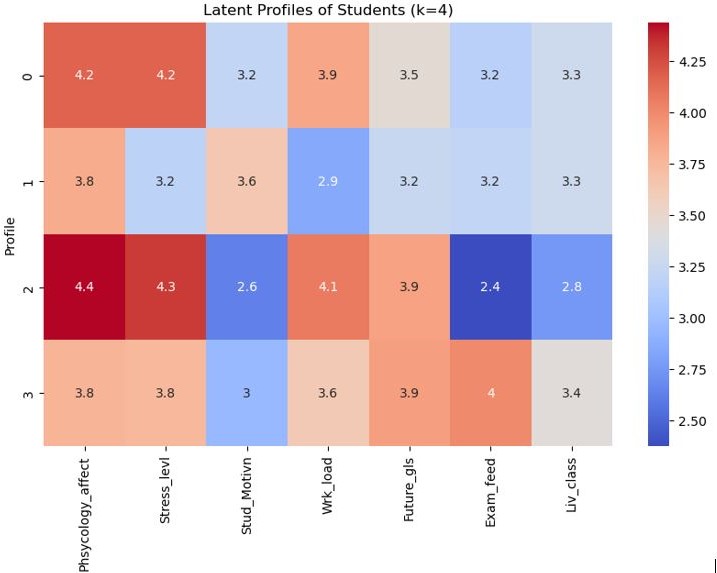}
 \caption{Latent student profiles based on most relevant variables}
    \label{heatmap}
\vspace{-1pt}
\end{figure}
\autoref{heatmap} illustrates that the four student profiles vary in terms of cognitive presence (future goals, motivation), social presence (psychological affect, stress), and teaching presence (exam feedback, live classes). Profile 1 students represent a strong balance across these presences: they show manageable stress, higher motivation, and consistent engagement, which aligns with optimal learning in blended settings. Profile 0 students exhibit high social presence pressures (stress and psychological effect) but weaker cognitive presence (lower motivation), suggesting that without emotional regulation, their engagement in learning is hindered. Profile 2 represents the most at-risk group, with high workload and stress (social presence overload) but very low cognitive and teaching presence (motivation and feedback), reflecting a breakdown in the CoI balance that undermines academic performance. Finally, Profile 3 shows moderate levels across dimensions, with relatively strong cognitive presence (future goals) but average social and teaching presence, placing them between the high-performing and at-risk groups. 

The LPA results suggest that students are not experiencing the blended classroom uniformly, instead, they are clustered into distinct groups with different balances of teaching, cognitive, and social presence. This raises an important theoretical implication: CoI framework, while valuable, may not fully capture the heterogeneity of learners in blended settings. For some groups, cognitive presence (e.g., motivation, future goals) is the main driver of performance, while for others, teaching presence (e.g., live classes, feedback) or even stress regulation plays a larger role. This implies that a single, one-size-fits-all application of CoI might overlook subgroup dynamics and that blended learning effectiveness may hinge on tailoring interventions to the needs of specific learner profiles. This suggest that we need to either adapt CoI for the variances in the blended learning setting or combine it with other frameworks for a more holistic outlook.

\subsection{Summary of key findings}
Overall, our empirical analysis demonstrated that teaching and social presences are critical elements for the success of the students cognitive engagement and learning experience in blended learning contexts. Summarized below are several key findings that we identified.
\begin{itemize}
    \item \textbf{Variations in academic engagement patterns}: The exploratory analysis of the survey data uncovered some discrepancies in learning habits and engagements (Section~\ref{Profiles}). For example, we found that female students across all age groups spend more time in learning compared to their male colleagues ($p<0.01$, t-test; \autoref{studytime}) and their overall academic outcomes were less negatively affected (\autoref{gender_vs_acper}). Linked with this, blended learning adjustments were also found to have very limited or no impact on senior and more mature postgraduate students ($>35$ years) compared to relatively younger peers ($<35$ years). These findings are well aligned with similar outcomes reported in several other closely related studies including the works of~\cite{harper2024face} and ~\cite{richardson2012psychological}.
    \item \textbf{Preference of synchronous learning activities}: Our study's results (Sections \ref{Sat_with_LT-methods1} and \ref{HypoTesting}) have shown that live online classes were more popular among students as indicated through their high satisfaction with this learning mode followed by recorded lectures (which may include a lower social interaction depending on the lecturer's teaching style). This could be attributed to the engagement and interaction inherent in these synchronous sessions.  From CoI framework perspective, live synchronous sessions can be considered as a strong component of teaching presence and our findings strongly suggest that the more effective this facilitation is, the better students perform. This also highlights that learning activities combining teaching and social presences are more effective and preferred by students. The result agrees with related literature which highlights that postgraduate students show greater engagement and satisfaction with synchronous learning activities~\citep{heilporn2021examination, fabriz2021impact}. 
  \item \textbf{Social interactions enhance learning experiences}: our empirical analysis and their projection to the CoI framework have also demonstrated that embedding social presences in the teaching delivery improves students’ cognitive engagement and promotes learning in blended learning contexts (Sections \ref{Sat_with_LT-methods1} and \ref{HypoTesting}). For instance, students have expressed highest satifaction with online live classes (\autoref{sat_with_LT-methods}, \autoref{fig:sat_with_LT-methods}) which was found to be highly correlation with learning performance ($p<0.01$, Spearman test, \autoref{SpearmanTest}), possibly due to its potential to provide a higher social interaction compared to flipped learning methods such as recorded lectures. The relevant literature points out how the lack of social interactions with the teaching and learning community (teachers, fellow students, etc) can undermine deep learning and cognitive engagement in blended learning context~\citep{fabriz2021impact,van2024fostering}. This could be attributed to the fact that social presence  builds a community where students feel connected, sense of belonging, and can express themselves.
\item \textbf{Psychological factors undermine academic success}: Through the analysis of the impact of various learning experience factors on academic performance, we found that all investigated mental factors, such as stress fluctuations, perceived workload, and motivational dynamics, are negatively correlated with students’ academic attainment ($p<0.01$, Spearman test, \autoref{SpearmanTest}). The undermining influences of these emotional challenges on learning and student outcomes are also well-established in related literature studies~\citep{harper2024face, richardson2012psychological,aristovnik2020impacts}.
\item \textbf{Student satisfaction as a driver of academic success}: the dependence tests between academic achievement and various BL factors have also found the highest consistent differences in learner satisfaction and agreements -- on learning methods, mental factors, and staff support -- between the two extreme groups, students with ‘improved’ and ‘worsened’ academic outcomes ($p<0.01$, MWU test, Tables \ref{tab:WMU-live_class}-\ref{tab:stress-vs-ap}). This means that the median satisfaction for academically successful students was consistently and statistically higher compared to underperforming learners. This association between satisfaction and learning  performance in blended learning settings is broadly acknowledged in the related literature ~\citep{kok2025mediating,aristovnik2020impacts}.
\item \textbf{Effective teaching support sustains academic success}: From the analysis on the perceived academic support for students (Section~\ref{StaffSupport}), we found that more than 60\% to 80\% of the surveyed students -- whose performance is either improved or maintained (\autoref{fig:agree_with_sup-types}), felt supported with their assessments by academic staff. This dependence of the academic support and achievement was also further confirmed with the statistical analysis presented in Section~\ref{HypoTesting}. The evidence shows that strong teaching presence through academic feedback and support boosts learner cognitive engagement in blended learning environments. This aligns with existing literature that highlight how teacher support enhances student engagement in higher education~\citep{prananto2025perceived}.
\end{itemize}

\section{Conclusion}
This paper investigated analysis of the relationships between postgraduate students’ academic performance and key dimensions of their learning experience—including perceived effectiveness of blended learning approaches, fluctuations in academic demands (e.g. workload) and emotional well-being (e.g., stress, psychological effect, motivation), and acknowledged adequacy of academic support and evaluative feedback—within blended learning contexts. Drawing on a quantitative survey data from 255 postgraduate students interpreted through the Community of Inquiry framework, our findings underscore the importance of fostering both teaching and social presence, alongside providing responsive academic support through effective feedback mechanisms. Importantly, the research findings highlight that the dynamics of academic load and student mental wellbeing can significantly hinder academic outcomes. Our data-driven analysis also revealed that fostering both instructional and interpersonal presence plays a pivotal role in enhancing students’ cognitive engagement and overall learning outcomes within blended learning environments. These insights call for a more nuanced and empathetic approach to blended learning design, one that balances technology-enabled pedagogy with social and human dimensions of learning. Finally, validated with the CoI framework, this study provides useful educational insights for educators to purposefully design blended learning modules and courses that are pedagogically effective but also socially enriching and emotionally supportive.

 \vspace{-5pt}
\paragraph*{Data availability:}The dataset created and used for the analysis of this study is available from the  authors on reasonable request. 
 \vspace{-10pt}
\paragraph*{AI use:} All writing, analysis, and interpretations presented in this paper were carried out by the authors. Artificial intelligence tools (Microsoft Copilot) were sometimes used to support literature collection and language improvements during the writing.
 \vspace{-10pt}
\paragraph*{Ethics statement:} The survey on which the analysis of the article is based was conducted as part of a research project that received formal institutional ethics approval. All data were collected and analysed anonymously, and no individual participant is identified in the manuscript.

\end{document}